\documentstyle[prl,aps]{revtex}
\input psfig
\begin{document}
\newcommand {\be}{\begin{equation}}
\newcommand {\ee}{\end{equation}}
\newcommand {\bea}{\begin{eqnarray}}
\newcommand {\eea}{\end{eqnarray}}
\newcommand {\nn}{\nonumber}
\newcommand{\ket}[1]{\mbox{$ | #1 \rangle $}}
\newcommand{\matel}[3]{\mbox{$ \langle #1 | #2 | #3 \rangle $}}
\newcommand{\expect}[1]{\mbox{$ \langle #1 \rangle $}}
\draft
\twocolumn[\hsize\textwidth\columnwidth\hsize\csname @twocolumnfalse\endcsname
%
%
%

\title{Generalization of the Luttinger Theorem for Fermionic Ladder Systems}

\author{Patrick Gagliardini, Stephan Haas, and T. M. Rice}
\address{Theoretische Physik,
ETH-H\"onggerberg, CH-8093 Zurich, Switzerland}

\date{\today}
\maketitle

\begin{abstract}
We apply a generalized version of the Lieb-Schultz-Mattis
Theorem to fermionic ladder systems to  show the existence of a low-lying
excited state (except for some special fillings).
This can be regarded as a non-perturbative proof for  the conservation
under interaction of
the sum of the Fermi wave vectors of the individual channels,
corresponding to a generalized version of the
Luttinger Theorem to  fermionic ladder systems.
We conclude by noticing that the Lieb-Schultz-Mattis Theorem is not applicable in
this form to show the existence of low-lying excitations
in the limit that the number of legs goes  to infinity, e.g. in the limit of a
2D plane.
\end{abstract}

\vskip2pc]
\narrowtext

In their paper of 1961 \cite{Lieb} Lieb, Schultz and Mattis 
(LSM) showed
as an exact result that an antiferromagnetic Heisenberg spin-$\frac{1}{2}$
chain of length L (even)
with periodic boundary conditions has gapless excitations in the thermodynamic
limit.
In one dimension, $S=\frac{1}{2}$ spins can be mapped into spinless
fermions by the Jordan-Wigner transformation. As a result, the
LSM Theorem should also be applicable to one-dimensional fermion
problems.
This was demonstrated recently by Yamanaka, Oshikawa and Affleck  \cite{Yama},
who considered a translationally invariant Hamiltonian with short-range hopping, 
which  conserves the number of up- and down-spin separately 
and is invariant under parity or time reversal.
Under these assumptions, the
generalized LSM Theorem can be stated as follows:
 in a chain of length L with
periodic boundary conditions an excited state with energy $O(\frac{1}{L})$
above the ground state exists at momentum $2\pi\nu_\sigma$, if the density $\nu_{\sigma}$ of fermions with spin
$\sigma$  per unit cell is not an integer and the ground state is not degenerate.

In a non-interacting one-dimensional system, the presence of gapless
excitations at momentum $2k_F$ is a consequence of the existence of Fermi
points at $\pm k_F$.
Luttinger showed perturbatively the existence of a meaningful concept of Fermi
surface in an interacting system, identifying it as the surface located at the
singular wave vectors of  the momentum distribution  function $n({\bf k})$, which enclose the same volume as
in the non-interacting system \cite{Lutt}.
Luttinger's proof applies to systems belonging to the Landau-Fermi liquid universality
class, and hence not directly to 1D problems.
The existence of gapless excitations at $2k_F=2\pi\nu_\sigma$ in an
interacting system can be seen as a generalized version of the Luttinger
Theorem in one dimension.

The application of the LSM Theorem  to spinful electrons
interacting on a ladder allows us to obtain a generalized Luttinger
Theorem for such systems.
In this geometry, we consider  the Hamiltonian:
\begin{eqnarray} 
  H 
&=&  \ -t \sum_{\alpha=1}^{\lambda}\sum_{j,\sigma}  {c}^{\dagger
  }_{\alpha,j+1,\sigma} { c}_{\alpha,j,\sigma} + H.c.\nonumber\\ 
&-&   t' \sum_{\alpha=1}^{\lambda-1}\sum_{j,\sigma}  { c}^{\dagger
  }_{\alpha+1,j,\sigma} { c}_{\alpha,j,\sigma} + H.c. + H_{int} =  H_t + H_{int}\nonumber
\end{eqnarray}
where $c_{\alpha,j,\sigma}$ is the annihilation operator for a fermion on the chain $
\alpha$, rung j, and with spin $\sigma$. $t$ and $t'$ are the hopping
amplitudes along the chains and along the rungs respectively.  Periodic boundary conditions are used along each
chain: $c_{\alpha,j+L,\sigma}=c_{\alpha,j,\sigma}$, $\alpha=1,..,\lambda$ (see
Fig 1). 

\begin{figure}
\centerline{\psfig{figure=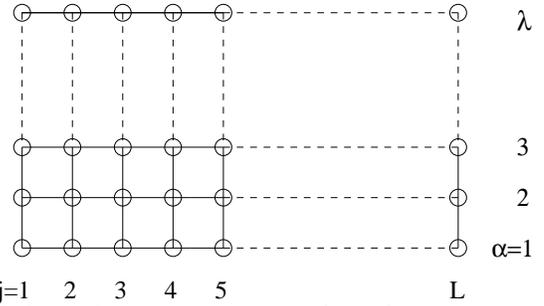,height=4.0cm,angle=0}}
\caption{
A $\lambda$-leg ladder. The lattice points are denoted by small circles.
}
\end{figure}

We assume that $H_{int}$ involves the local fermion densities only, that
$H$ preserves the number of up- and down-spin separately, and that it is
translationally 
and parity invariant. Here translation and parity operators are defined by $T { c}_{\alpha,j,\sigma }
  T^{-1} ={ c}_{\alpha,j+1,\sigma }$, and $P { c}_{\alpha,j,\sigma } P^{-1} ={ c}_{\alpha,-j,\sigma }={
  c}_{\alpha,L-j,\sigma }$, respectively.

Defining the twist operator $U_\sigma$, $\sigma=\uparrow,\downarrow$, by
\begin{equation}
{ U}_{\sigma} = \exp(2 \pi i\sum_{\alpha,j} \frac{j}{L} {
  n}_{\alpha,j,\sigma}),
\label{Thetw}
\end{equation}
 we obtain ${ U}^{-1}_{\sigma}{ c}^{\dagger }_{\alpha,j,\sigma} 
{ U}_{\sigma}={ e}^{-i\frac{2\pi}{L}j }{ c}^{\dagger }_{\alpha,j,\sigma}$, and
thus $H_{int}$ and the kinetic term for the hopping along the rungs are
invariant under $U_\sigma$.
Letting $\ket{\phi_0}$ be the ground state of the system, the excitation
energy of the twisted state $\ket{\phi_1}=U_\sigma \ket{\phi_0}$ is
$E_1 = \matel{\phi_0}{{ U}^{-1}_{\sigma}H{ U}_{\sigma} - H}{\phi_0}
=-t({ e}^{-i\frac{2\pi}{L} } - 1)\sum_{\alpha,j}  
\expect{ { c}^{\dagger }_{\alpha,j+1,\sigma} { c}_{\alpha,j,\sigma}}  +
H.c.$ .
Expanding the exponential and assuming a parity invariant ground state, the
term $O(1)$, involving expectation values like $\expect{ i \sum_{j} { c}^{\dagger }_{\alpha,j+1,\sigma} { c}_{\alpha,j,\sigma}-
{ c}^{\dagger }_{\alpha,j,\sigma} { c}_{\alpha,j+1,\sigma}} $, vanishes, and
we thus  obtain: $E_1 = O(\frac{\lambda}{L})$.

From $T U_\sigma T^{-1} =e^{- 2\pi i\lambda   { \nu}_{\sigma}}U_{\sigma}$ we
deduce that the crystal momentum $P$ of $\ket{\phi_1}$, defined by
$T=e^{-iP}$, is  $2\pi\lambda   {\sf \nu}_{\sigma}$, relative to the ground
state. 
This proves that the twisted state $\ket{\phi_1}$ is orthogonal to
$\ket{\phi_0}$ if $\lambda\nu_\sigma$ is not an integer. In this case, the
existence  of at least
one low-lying  excited state is assured variationally, as long as  the ground
state is unique.

Thus, the LSM Theorem can be generalized to interacting fermions on a $\lambda$-leg
ladder as follows:  under the above assumptions for the Hamiltonian H and
its ground state, gapless excitations exist in the thermodynamic limit at momentum  $2\pi\lambda   {\sf
  \nu}_{\sigma}$,  if
 $ \lambda   {\sf \nu}_{\sigma}$ is not an
  integer.

We notice that this method, being essentially topological in nature,  does not depend on
the details of the
Hamiltonian, but only on general properties. For example, if a
next-nearest neighbor hopping term is introduced, it  gives a contribution of
order $O(\frac{\lambda}{L})$ to the excitation energy of the twisted state, and the
proof applies as well.

Following Yamanaka et al. \cite{Yama}, we define charge and spin twist operators by
\begin{equation}
U_c=U_\uparrow U_\downarrow, \qquad U_s=U_\uparrow U^{-1}_\downarrow .
\end{equation}
Under their action,  the electron spin operators  ${\sf {\bf S}}_{\alpha,j}={ c}^{\dagger}_{\alpha,j,\sigma}\frac{ {\bf
    \tau}_{\sigma,\sigma'}}{2}{ c}_{\alpha,j,\sigma'}$ 
transform like
\begin{equation}
U^{-1}_c { {\bf S}}_{\alpha,j}U_c = { {\bf S}}_{\alpha,j} \qquad 
U^{-1}_s { {\bf S}}_{\alpha,j}U_s = { {\bf R}}_3 (\frac{4\pi }{L}j){ {\bf S}}_{\alpha,j},
\end{equation}
where $ {\bf
  R}_3(\alpha)$ is the rotation about the third axis with angle $\alpha$.
In this sense, the operators $U_c$ and $U_s$ can be interpreted as  twist operators creating a
charge and a spin excitation respectively.
From $TU_cT^{-1}=U_ce^{- 2\pi i \lambda  {\sf \nu}}$, where $\nu =
\nu_\uparrow + \nu_\downarrow$ and
$TU_sT^{-1}=U_se^{- 2\pi i\lambda (  {\sf \nu}_\uparrow -{\sf
    \nu}_\downarrow)}$, we see that  gapless charge (spin)
excitations exist if $\lambda\nu$ ($\lambda ( {\sf \nu}_\uparrow -{\sf
  \nu}_\downarrow)$) is not an integer.
If no magnetization is present $\nu_\uparrow=\nu_\downarrow=\frac{\nu}{2} $,
 a statement about the existence of gapless spin excitations cannot  be made by
the above method.

At half-filling ($\nu_\sigma=\frac{1}{2}$), a gapless excitation is shown to exist
(if  the above assumptions hold) in a ladder with an odd number of
legs. However, in all cases at this filling a charge gap can open (due to
relevant Umklapp processes). 
Furthermore, unlike the case of a single chain, in a $\lambda$-leg ladder a charge gap can
open (without breaking the translational symmetry of the ground state) even at
rational fillings, i.e. $\nu=\frac{m}{\lambda}$, with $m$ integer.
These conclusions are consistent with perturbative and numerical studies
\cite{Lin}.

For the Hubbard model, the application of the generalized
LSM Theorem leads to conclusions which are independent on the
sign of U.
However, the physics of the attractive and of the repulsive  Hubbard model are
very different: on a chain, for $U<0$, the model scales to strong-coupling and
falls in the universality class of the Luther-Emery Liquid \cite{Emelu}, with gapped spin
excitations, whereas for $U>0$ it scales to the Tomonaga-Luttinger fixed point
 \cite{Tomo}, 
with gapless charge and spin excitations (away from half-filling).
(For reviews of interacting fermion systems in one dimension and bosonization
techniques, see e.g. \cite{Soly}, \cite{Hald}, \cite{Emery} and \cite{Affl}).
Hence, the presence of gapless excitations on a Hubbard chain away from
half-filling, without characterization of their nature, as obtained with the
LSM Theorem using $U_\sigma$, is in agreement with perturbative results.
However, the differences in the spin excitation spectrum  between the various
regions of the parametric space cannot be observed with the LSM Theorem, since
the momentum of the  excitations created by $U_s$ vanishes (when no
magnetization is present).
To get further insight, we apply the LSM Theorem 
to the strong coupling limits ($U \longrightarrow \pm \infty$) of the
Hubbard model  separately.

In the attractive case, for large $|U|$, the electrons form Cooper pairs with
opposite spins, occupying  equally spaced
lattice points, and
the Hamiltonian reduces (up to constant terms) to one for hard-core bosons
 on a lattice with a repulsive next-neighbor interaction 
\cite{Efetov}.
An appropriate twist operator can be defined to show that
 gapless charge excitations exist at momentum $\pi
 \nu$, if $ \nu$ is not an even integer and the ground state is
not degenerate. 
This shows that the low-lying excitations at momentum $2\pi\nu_{\sigma}$,
whose existence was shown above at finite $U$ without specifying their nature,
are actually charge excitations, in the attractive case.
Spin excitations require the breaking of a Cooper pair  and cost an energy amount of order $|U|$.

In the case $U>0$, the strong on-site repulsion favors single occupation of
the lattice sites by the electrons, and thus the Hubbard model at strong-coupling
scales to the t-J Hamiltonian, for which we can show as well that 
$E_1=O(\frac{1}{L})$, noticing that the twist operator acts on the electron
spin operator as one would expect: $U^{-1}_\sigma{\sf {\bf S}}_{j}U_\sigma={\sf {\bf R}_3}(\pm \frac{2\pi}{L}j){\sf {\bf
    S}}_{j}$, $\sigma=\uparrow,\downarrow$, and that it commutes
with the projector $ P_s=\prod_{j} ( 1 - { n}_{j \uparrow}{
  n}_{j \downarrow})$, which  prohibits double occupancy.

Recently, it was proposed  that the introduction of a next-nearest-neighbor
hopping term in the 1D t-J  model could lead to a breakdown of the Luttinger
liquid behavior \cite{Eder}. In particular, numerical calculations have suggested the existence of 
a low-doping phase with Fermi momentum $k_F = \frac{\pi}{2}\delta=\frac{\pi}{2}(1-\nu)$ (determined
thus by the density of holes), and a high-doping phase with $k_F=\pi \nu$
(determined thus by the density of spinless fermions); neither of these $k_F$
is compatible 
with a Luttinger liquid.
As pointed out above, the generalized LSM Theorem  shows the existence of  gapless
excitations at momentum $2k_F=\pi\nu$  even after the introduction of a
next-nearest hopping term, unless a
degeneracy in the ground state occurs. 
The above high-doping 
phase  is not consistent
with  our $2k_F$-excitation. The origin of this ``anomalous`` $k_F$ could
lie in ferromagnetic correlations. 

The generalization to the case of a $\lambda$-leg t-J ladder by means of
(\ref{Thetw}) is immediate: a low-energy excitation at momentum $2\pi\lambda\nu_\sigma$
is explicitly shown to exist in a $\lambda$-leg  t-J ladder, if 
$\lambda\nu_\sigma$ is not an integer.
At half-filling, the t-J model reduces to a Heisenberg (spin-only) Hamiltonian,
and thus the  low-lying excited states for $\lambda$ odd have the character
of spin excitations, as can be
shown with the specific
twist operator $U=e^{ 2\pi i \sum_{\alpha,j} \frac{j}{L}{\sf {\bf
     S}^3}_{\alpha,j}}$, introduced by Affleck \cite{Affleck}.
Using  the known property of spin-$\frac{1}{2}$
representations $e^{2\pi i S^3_{\alpha,j}}=-1$, and assuming a non-degenerate
singlet ground state in the thermodynamic limit, 
 one  gets $TUT^{-1}=(-1)^{\lambda}U $ (acting on the ground state), and
thus the relative momentum of the twisted state is $\pi$
(0) if $\lambda$ is odd (even). 



We next show how the identification of the band structure of a free
Hamiltonian defined on ladder allows us to visualize the
simple nature of $U_\sigma$ in the non-interacting model.
Defining  new Fermi operators $\tilde{{ c}}_{\mu,j,\sigma}=T_{\mu,\alpha}{
  c}_{\alpha,j,\sigma}$, $\mu=1,..,\lambda$, with an appropriate $ SO(\lambda)$ matrix $T$, a decoupling of the chains in the free 
kinetic Hamiltonian can be achieved. Fourier transforming on each channel ${ c}_{\mu,k,\sigma} =
\frac{1}{\sqrt{L}}\sum_{j} e^{ ikj}{ c}_{\mu,j,\sigma}$, we get: 
\begin{eqnarray} 
 H_t 
&=&  \ -t \sum_{\mu=1}^{\lambda}\sum_{j,\sigma}  { c}^{\dagger
 }_{\mu,j+1,\sigma} { c}_{\mu,j,\sigma} + H.c.
- t'\sum_{\mu=1}^{\lambda}\sum_{j,\sigma} \Lambda_\mu n_{\mu,j,\sigma}
\nonumber\\
&=&\sum_{\mu=1}^{\lambda}\sum_{k,\sigma}(-2t\cos(k)
-t'\Lambda_\mu)n_{\mu,k,\sigma} 
\end{eqnarray}
 where $\Lambda_\mu=2\cos( \pi \frac{\mu}{\lambda + 1})$.
The  channels $\mu$ can be classified by their 
parity under the reflection $R$ about the center leg (about the center axis), in
the case of odd (even) $\lambda$: channels with $\mu$
even  are odd under $R$ and those with $\mu$ odd are even. Furthermore, comparing
$\epsilon_\mu(k)=-2t\cos(k)
-t'\Lambda_\mu$ with the dispersion relation for the 2D model, we can associate with the band $\mu$ the transverse momentum
$k_{\perp,\mu}=\pi \frac{\mu}{\lambda + 1}$, and form a 2D momentum ${\bf k}=(k,k_{\perp,\mu})$.
We can now consider  twist operators for each 
channel, ${ U}_{\mu,\sigma} = \exp(2 \pi i\sum_{j} \frac{j}{L} {
  n}_{\mu,j,\sigma})$, with $E_{\mu,1} =\matel{ \phi_0}{  {
    U}^{-1}_{\mu,\sigma}H_t { U}_{\mu, \sigma} - H_t}{\phi_0}=O(\frac{1}{L})$, and $
TU_{\mu,\sigma}T^{-1}=U_{\mu,\sigma}e^{- 2\pi i   {\sf \nu}_{\mu,\sigma}}$,
where $\nu_{\mu,\sigma}=\frac{1}{L} \sum_j n_{\mu,j,\sigma}$.
Thus,  if $\nu_{\mu,\sigma}$ is not an integer, i.e. the
band $\mu$ is partially filled, a gapless excited state with momentum
$2\pi\nu_{\mu,\sigma}$ exists.
The action of $U_{\mu,\sigma}$ in momentum space is:
\begin{displaymath}
U^{-1}_{\mu,\sigma}{ c}^{\dagger}_{\mu',k,\sigma'}U_{\mu,\sigma}=
\left\{
\begin{array}{ll}
{
  c}^{\dagger}_{\mu',k+\frac{2\pi}{L},\sigma'} & \textrm{if $\sigma=\sigma'$
  and $\mu=\mu'$} \\
{
  c}^{\dagger}_{\mu',k,\sigma'} & \textrm{otherwise}
\end{array} \right.
\end{displaymath}
Acting on the non-interacting ground state, $U_{\mu,\sigma}$ creates a new eigenstate of
the non-interacting system with the net effect of  moving  one electron with spin
$\sigma$ and transverse momentum  $k_{\perp,\mu}$ from the left to the
right Fermi point in the channel $\mu$.
Because $\sum_\mu
n_{\mu,j,\sigma}=\sum_\alpha n_{\alpha,j,\sigma}$, we have $U_\sigma
=\prod_\mu U_{\mu,\sigma}$, and under the action of
$U_\sigma$, all the fermions with spin $\sigma$ are shifted one unit to the
right. Thus, the excitation created by $U_\sigma$ in the non-interacting system
corresponds to a rigid translation of the Fermi surface, moving an electron from one Fermi point to the other in each
channel, and has a
total momentum $\sum_\mu 2k_{F,\mu}$. Notice that, in the limit of a large
number of legs,  the volume enclosed by the
Fermi surface can be expressed as $V_{FS} = \frac{\pi}{\lambda + 1}\sum_\mu 2k_{F,\mu}$. 
\begin{figure}
\centerline{\psfig{figure=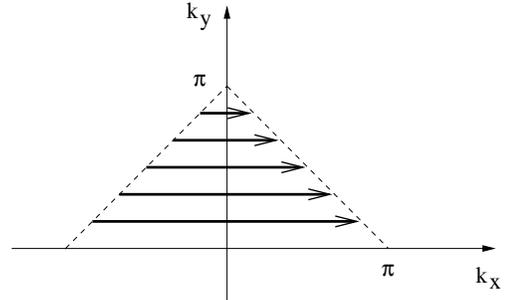,height=4.0cm,angle=0}}
\caption{
Global excitation created by $U_\sigma$ in the free system. The single
excitations at the different transverse momenta are shown. The dashed lines 
join the Fermi points of the channels in a system at half-filling.
}
\end{figure}
As seen before, the global excitation created by $U_\sigma$ has a vanishing
energy in the thermodynamic limit even for the interacting system, but this is
in general not the case for those created by a single $U_{\mu,\sigma}$.
Hence, in the individual channels, the Fermi wave vectors are not conserved
under the interaction, and Fermi points must not necessarily
 exist.
 However, the  conserved quantity in the interacting system is the sum of the
 Fermi wave vectors of all the channels, $\sum_\mu 2k_{F,\mu}$, in the sense that 
the excitation created by $U_\sigma$ remains gapless.
This can be regarded as a generalized Luttinger's Theorem for ladder systems.

To illustrate this point, we consider a 3-chain t-J ladder.
The non-interacting band structure has the form of 2 even-parity bands (bonding
and anti-bonding) and an odd-parity channel (non-bonding), whose Fermi momenta
at half-filling are $\frac{3}{4}\pi$, $\frac{\pi}{4}$, and $\frac{\pi}{2}$
respectively.
\begin{figure}
\centerline{\psfig{figure=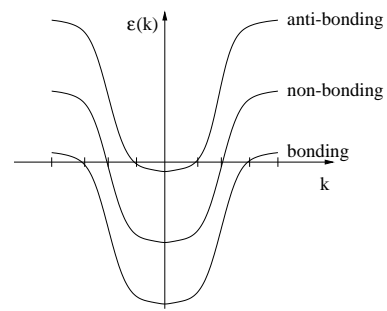,height=4.0cm,angle=0}}
\caption{
Free bands for a 3-leg ladder.
}
\end{figure}
Numerical calculations based on exact diagonalization
of small clusters \cite{Haas} show that the ground state for the undoped 3-leg
t-J ladder and the ground state for the same system with one hole have
opposite parities. A parity -1 is therefore associated with a single hole, and the
interpretation, obtained from the investigation of two and more holes, is that
below  a critical hole density $\delta_c$, the doped holes enter
the odd parity channel and form a Luttinger Liquid, while the two even parity
channels combine to form an insulating spin liquid.
Hence, for $\delta=1-\nu < \delta_c$, one electron per rung is assigned to each
of the two gapped insulating even-parity channels, $\nu_b + \nu_a = 2$,
resulting in  $2k_F=2\pi$ for the spin liquid, whereas the remaining $1 -
3\delta$ electrons per rung are assigned to the odd-parity channel,
$\nu_{nb}=1-3\delta$, with $2k_F=\pi(1-3\delta)$ for the Luttinger liquid.
Gapless excitations at $\pi(1-3\delta)$ are consistent with our results:
$\pi(1-3\delta)\equiv \pi(1-3\delta) + 2\pi=3\pi(1-\delta)=
3\pi\nu$.
Thus, the combination with the two even-parity channels in a gapped spin liquid
 leads to the generalized
Luttinger Theorem for a 3-leg ladder: $\sum_\mu 2k_{F,\mu} =3\pi\nu$. 

We have seen that the band structure of a non-interacting fermionic ladder system results in
momentum states aligned 
at different transverse momenta parallel to the $k_x$-axis. We can show that
an equivalent construction is possible in other directions in momentum space,
starting from ladder structures in other directions than that of the lattice axis.
In this case, the generalized LSM Theorem can be applied to
show the existence of  a low-energy excitation on a $\lambda$-leg ladder at momentum
$2\pi\lambda\nu_\sigma$ in each direction of periodicity of the 2D cubic
lattice, if $\lambda\nu_\sigma$ is not an integer.
Let us consider a ladder $G$ with $ \lambda$ legs defined by a vector $ {\bf T}
=(p,q) $, p,q $\in  \mathbf{N} $ coprimes, which joins 2 lattice points. Let $\bf n$
be a unit vector in the direction of the ladder, and $\Lambda$ be the
quantization length along the ladder direction.
\begin{figure}
\centerline{\psfig{figure=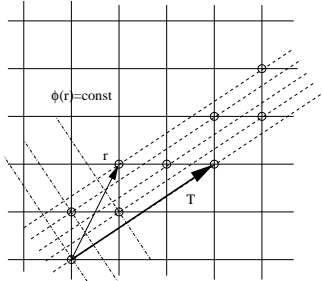,height=4.0cm,angle=0}}
\caption{
A $\lambda$=5-leg ladder in direction ${\bf T}$. The ladder points are defined
by the intersection of the diagonal lines with the lattice axis and are
denoted by small circles.
}
\end{figure}
A generalized twist operator is $U_\sigma = e^{ i \sum_{\bf r \in G} \varphi ({\bf r} )n_{\bf r, \sigma}}$, where $\varphi ({\bf
 r})=\frac{2\pi}{\Lambda}{\bf r} \cdot {\bf n}$.
By means of an inversion center of $G$, a parity operator is defined, and an 
excitation energy of $O(\frac{\lambda}{L})$ for the twisted state can be  shown as before.


What can be said in this context about the whole 2D plane?
To get insight into the full 2D problem letting the number
of the legs going to infinity, we consider $
H=-t\sum_{{\bf r},\tau,\sigma} c^\dagger_{\bf r + \tau,\sigma}c_{\bf r,\sigma}
+ H_{int}$
 where ${\bf r}=(j,\alpha)$, $j=1,..,L$, $\alpha=1,..,\lambda$, and the sum over
 $\tau$ extends over next-neighbors: $\tau = \pm {\bf e}_x$, $\pm {\bf e}_y$. Periodic boundary conditions are now used for both directions: $c_{{\bf r} + L{\bf
    e}_x + \lambda{\bf e}_y,\sigma}=c_{\bf r,\sigma}$.
Obviously, we have again  $E_1=\expect{U^{-1}_\sigma H U_\sigma -
H}=O(\frac{\lambda}{L})$.
We see that the thermodynamic limit $\lambda L \longrightarrow \infty$ of $E_1$
does not exist. For example, letting $\lambda,L \longrightarrow \infty$ with
$\frac{\lambda}{L} \longrightarrow 1$, we have $E_1 \sim $ const.
This is in agreement with the action of $U_\sigma$ on the ground state for
the non-interacting model.
In fact, Fourier transforming in a 2D momentum space $c_{ {\bf k},
  \sigma}=\frac{1}{\sqrt{\lambda L}} \sum_{ {\bf r} }  e^{i{\bf k} \cdot {\bf
    r} } c_{ {\bf r}, \sigma}$, we have $U^{-1}_{\sigma}c^\dagger_{ {\bf
    k},\sigma}U_\sigma=c^\dagger_{ {\bf k + q},\sigma}$, with ${\bf
q}=\frac{2\pi}{L}{\bf e}_x$.
Hence, the whole Fermi surface is shifted, and even if each
single excitation of an electron at a given transverse momentum costs a
vanishing energy, the number of such
excitations becomes large, as $\lambda$ increases.
We conclude that the LSM Theorem in this form does not allow
to show the existence of a low-energy excitation in the whole 2D plane.

In summary, on a ladder with an arbitrary fixed number of
legs $\lambda$, a low-lying excitation exists at momenta
$2\pi\lambda\nu_\sigma$ (in each direction of periodicity of the 2D square
lattice), if $\lambda\nu_\sigma$ is not an integer. 
The existence of these gapless excitations can be regarded as a statement of
the conservation under interaction of the sum of the Fermi wave vectors of the
individual channels, corresponding to a generalized Luttinger Theorem for
ladder systems.

We wish to thank M. Sigrist for useful discussions, and acknowledge the Swiss
National Science Foundation for financial support.

%
%

\end{document}